\documentclass[prl,aps,superscriptaddress,longbibliography,twocolumn,nofootinbib]{revtex4-2}
\usepackage[table]{xcolor}
\usepackage[colorlinks=false]{hyperref}
\usepackage[T1]{fontenc}
\usepackage[utf8]{inputenc}
\usepackage{color}
\usepackage{siunitx}
\usepackage{bbm} 
\usepackage[english]{babel}
\usepackage{multirow}
\usepackage{amsfonts,amsmath,amssymb,stmaryrd}
\usepackage{braket}
\usepackage{graphicx}
\usepackage{epsfig}
\usepackage{mathrsfs}
\usepackage{verbatim}
\usepackage{centernot}
\usepackage{ulem}
\usepackage{longtable}
\usepackage{nicefrac}
\usepackage[framemethod=tikz]{mdframed}
\usepackage[nolist]{acronym}
\usepackage{soul}
\usepackage{array}
\usepackage{booktabs}   
\usepackage{amsmath}    
\usepackage{amsfonts}   
\usepackage{xcolor}

\usepackage{cancel,ifthen}
\newcommand{\cmnt}[2][NoInPuT]{\ifthenelse{\equal{#1}{NoInPuT}}{}{{\color{red}\sout{#1}}} {\color{blue} #2}}

\usepackage{bm}	
\renewcommand{\vec}[1]{{\bm{#1}}}

\LTcapwidth=\textwidth

\begin{document}

\newcommand{\opd}[1]{\hat{\mathop{#1}}^\dagger}  
\newcommand{\op}[1]{\hat{\mathop{#1}}}           
\newcommand{\opp}[1]{\hat{\mathop{#1}}^+}        
\newcommand{\opm}[1]{\hat{\mathop{#1}}^-}      
\renewcommand{\i}{\text{i}}                   
\newcommand{\e}[1]{\mathrm{e}^{#1}}            

\normalem	

\title{Symmetry-Enforced Chiral Phonons in Altermagnets via Magnon-Phonon Coupling}

\author{Philipp Rieger}
\email[]{philipp.rieger@physics.uu.se}
\affiliation{Department of Physics and Astronomy, Uppsala University, P. O. Box 516, S-751 20 Uppsala, Sweden}

\author{Markus Weißenhofer}
\affiliation{Department of Physics and Astronomy, Uppsala University, P. O. Box 516, S-751 20 Uppsala, Sweden}

\author{Sergiy Mankovsky}
\affiliation{Institute  of  Materials  Science, Technical University of
Darmstadt, 64287  Darmstadt, Germany}
\affiliation{Department of Chemistry/Phys.\ Chemistry, LMU Munich,
   Butenandtstrasse 11, D-81377 Munich, Germany}

\author{Peter M. Oppeneer}
\affiliation{Department of Physics and Astronomy, Uppsala University, P. O. Box 516, S-751 20 Uppsala, Sweden}

\date{\today}

\begin{abstract}
    Chiral phonons are attractive for spintronics applications, however, their zero-field generation in conventional antiferromagnets is forbidden by combined parity and time-reversal ($\mathcal{PT}$) symmetry.  
    Here we demonstrate the emergence of chiral phonons in $\mathcal{PT}$-breaking altermagnetic systems at zero field arising from relativistic magnon-phonon coupling. 
    Focusing on the prototypical altermagnet CrSb, we utilize first-principles methods to calculate the hybridized magnon-polarons across the complete Brillouin zone. 
    We show that this coupling imprints an altermagnetic $g$-wave symmetry directly onto the phonon angular momentum. Furthermore, we demonstrate anomalous spin and phonon angular momentum Nernst responses arising from finite Berry curvatures.
    These findings establish that chiral lattice dynamics can arise in compensated magnetic ground states without requiring external fields, positioning bulk altermagnets as material candidates for zero-field spin caloritronics and chiral phononics.
\end{abstract}

\maketitle

\textit{Introduction---}
Lattice vibrations possessing angular momentum have emerged as a 
degree of freedom that can be detected, manipulated, and exploited 
in condensed matter physics. Following foundational theory \cite{Zhang_Niu_PRL, zhang_niu_2015_PhysRevLett.115.115502} this phonon angular momentum (PAM) has been detected in noncentrosymmetric systems via { circular dichroism \cite{Zhu_2018_doi:10.1126/science.aar2711}}, x-ray \cite{ueda_chiral_2023},  Raman \cite{Kyosuke_2023, ishito_truly_2023}, transport \cite{Kazuki_2024_PhysRevLett.132.056302, kim_chiral-phonon-activated_2023}, and torque  \cite{zhang_measurement_2025} measurements. 
These modes—termed \textit{axial}, or \textit{chiral} when the PAM pseudovector aligns with the { time-domain} propagation direction \cite{chiralnature_jurascheck_2025}—are associated with large phonon magnetic moments \cite{Juraschek_PhysRevMaterials.3.064405, Juraschek_PhysRevMaterials.1.014401, phon_mag_mom_Ren_PhysRevLett.127.186403, basini_terahertz_2024}, ultrafast demagnetization \cite{tauchert_polarized_2022, dornes_ultrafast_2019}, and the ultrafast reversal of magnetic order via phononic Barnett effects \cite{Luo_2023_doi:10.1126/science.adi9601, davies_phononic_2024}. 
In the steady state, PAM underlies anomalous transverse transport under thermal gradients: a universal PAM Hall effect arises even in nonrelativistic lattices without structural chirality \cite{Park_2020_doi:10.1021/acs.nanolett.0c03220}, while structural chirality enables a chiral-phonon-activated spin Seebeck effect \cite{kim_chiral-phonon-activated_2023}. 

From a symmetry perspective, bare phonons in centrosymmetric lattices describe purely linear, nonaxial vibrations \cite{yang_symmetry-guided_2026}. This constraint is lifted only by relativistic spin-lattice coupling (SLC). 
When the SLC hybridizes spin and lattice degrees of freedom in a $\mathcal{T}$-breaking magnetic environment, the resulting magnon-polarons \cite{berk_strongly_2019, luo_evidence_2023} can carry finite PAM \cite{holanda_detecting_2018}, as recently predicted in ferromagnets \cite{Markus_2025_PRL, Bonini2023_PhysRevLett.130.086701, Ma_2024_PhysRevLett.133.246604}. 
Simultaneously, the $\mathcal{T}$-breaking environment, transmitted to the lattice via SLC, permits finite Berry curvature. This drives anomalous thermal \cite{zhang_thermal_2019, Ma_2024_PhysRevLett.133.246604} and spin Nernst \cite{Markus_2025_PRL} effects.
However, both effects — finite PAM and the Berry curvatures that drive anomalous transverse transport — are forbidden at zero field in conventional collinear antiferromagnets,  as their ground state preserves combined space-inversion $\mathcal{P}$ and $\mathcal{T}$ symmetry. 
Since $\mathcal{P}$ and $\mathcal{T}$ impose opposite signs on the PAM at each $\vec{k}$-point, and Berry curvature is odd under $\mathcal{PT}$, both vanish identically throughout the Brillouin zone  (BZ).
Consequently, these zero-field quasiparticles remain nonaxial and topologically trivial.

Here, we show that altermagnets bypass this constraint. Despite maintaining a compensated collinear magnetic ground state, they break $\mathcal{PT}$ symmetry, yielding the spin-split electronic \cite{doi:10.7566/JPSJ.88.123702, Smejkal_PhysRevX.12.031042, krempasky_altermagnetic_2024, reimers_direct_2024, feng_anomalous_2022} and magnonic \cite{Smejkal_2023_PhysRevLett.131.256703, Cui_PhysRevB.108.L180401, weissenhofer_2024_PhysRevB.110.094427} bands that define the altermagnetic phase.
We show that this broken $\mathcal{PT}$ symmetry is not confined to the spin system.
We  demonstrate how relativistic SLC hybridizes altermagnetic magnons with phonons, imprinting the unconventional symmetry of the altermagnetic phase onto the phonon angular momentum distribution. 
Specifically, we present first-principles calculations demonstrating the zero-field emergence of chiral phonons in the altermagnet CrSb \cite{reimers_direct_2024, Ding2024, singh2025chiral,Li2025,Yang2025,Zhou2025}. 
This magnon-polaron formation induces finite Berry curvatures in the hybridized bands, driving anomalous spin Nernst transport and introducing relativistic modifications to the PAM Nernst response.

\textit{Atomistic spin-lattice coupling}--- We employ an atomistic framework to model the interaction between spin and lattice degrees of freedom. The total Hamiltonian $\hat{\mathcal{H}}$, comprising harmonic lattice dynamics, magnetic exchange, and SLC, reads:
\begin{align}
    \hat{\mathcal{H}}_\mathrm{} &= \sum_{i,l} \frac{(\hat{{P}}^\alpha_{il})^2}{2 M_l} 
    + \frac{1}{2} \sum_{ij,ll^\prime}  \Phi_{il,jl^\prime}^{\alpha\beta} \hat{u}_{il}^\alpha \hat{u}_{jl^\prime}^\beta \nonumber \\
    &+  \sum_{ij,nm}\frac{1}{S_n S_m}  J_{in,jm}^{\alpha\beta} \hat{S}_{in}^\alpha \hat{S}_{jm}^\beta \nonumber \\
    &+  \sum_{ijk, nml} \frac{1}{S_n S_m} J_{in,jm,kl}^{\alpha\beta\mu} \hat{S}_{in}^\alpha \hat{S}_{jm}^\beta \hat{u}_{kl}^\mu .
    \label{eq:SLCHamil}
\end{align}
The first two terms describe the harmonic lattice. Here, $M_l$, $\hat{P}_{il}^\alpha$, and $\hat{u}_{il}^\mu$ represent the mass, momentum, and displacement of basis atom $l$ in unit cell $i$. 
The real-space interatomic force constants (IFCs) $\Phi_{il,jl^\prime}^{\alpha\beta}$ yield the phonon frequencies $\omega_{\mathbf{k}\lambda}$ and eigenvectors $\vec{e}_{\mathbf{k}\lambda l}$ for branch $\lambda$ via the Fourier-transformed dynamical matrix $\mathcal{D}_{\mathbf{k},ll^\prime}$, by solving the eigenvalue problem $\sum_{l^\prime} \mathcal{D}_{\mathbf{k}, ll^\prime} \vec{e}_{\mathbf{k}\lambda l^\prime} = \omega_{\mathbf{k}\lambda}^2 \vec{e}_{\mathbf{k}\lambda l}$ in absence of relativistic effects \cite{Zhang_Niu_PRL}. ${\alpha,\beta,\mu}$ specify Cartesian coordinates, with Einstein summation implied. We calculate the IFCs using VASP \cite{Kresse1996} and phonopy \cite{phonopy-phono3py-JPCM,phonopy-phono3py-JPSJ, supplemental_material}. 

The third term captures spin dynamics via a generalized Heisenberg exchange $J_{in,jm}^{\alpha\beta}$, including isotropic exchange, magneto-crystalline anisotropy, and the antisymmetric Dzyaloshinskii-Moriya interaction (DMI). $\hat{S}^\alpha_{in}$ is the spin operator at magnetic site $n$, with magnitude $|\hat{\vec{S}}_{in}| = S_n$.

The final term introduces the SLC by accounting for the explicit dependence of the magnetic exchange interactions on atomic displacements. A first-order Taylor expansion around the equilibrium coordinates yields the SLC tensor $J_{in,jm,kl}^{\alpha\beta\mu} = \partial J_{in,jm}^{\alpha\beta} / \partial u_{kl}^\mu$ \cite{Hellsvik_PhysRevB.99.104302}. 
This derivative accounts for the full crystalline environment, capturing how the exchange between a magnetic pair $n,m$ is altered by the displacements of all neighboring atoms $l$, including the mediating nonmagnetic ligands. 
The magnetic parameters were calculated entirely from first principles within the fully relativistic KKR Green's function method \cite{EKM11,SPR-KKR8.5}. The exchange coupling tensor $J^{\alpha\beta}_{in,jm}$ was obtained using an approach based on the magnetic force theorem \cite{EM09}, which acts as a relativistic extension of the LKAG formula \cite{LKAG87}, while the SLC tensor $J^{\alpha\beta\mu}_{in,jm,kl}$ was calculated using an efficient scheme also based on the force theorem \cite{mankovsky_PhysRevLett.129.067202_2022, PhysRevB.107.144428, PhysRevB.107.115176}, giving access to the SLC parameters of any order with respect to perturbations due to atomic displacements.

Mapping localized spins to magnon operators ($\hat{b}_{\mathbf{k}n}^{(\dagger)}$) via the Holstein-Primakoff transformation \cite{PhysRev.58.1098} and atomic displacements to phonon operators ($\hat{a}_{\mathbf{k}\lambda}^{(\dagger)}$) yields the magnon-phonon Hamiltonian. For a multi-sublattice system collinearly aligned along the $z$-axis such as CrSb, this reads:
\begin{align}
    \hat{\mathcal{H}}_\mathrm{mp} &= \sum_{\vec{k},\lambda} \hbar\omega_{\vec{k}\lambda} \opd{a}_{\vec{k}\lambda} \op{a}_{\vec{k}\lambda}^{~} \nonumber \\
    &+ \sum_{\vec{k}, n,m} \left[ \mathcal{A}_{\vec{k}nm} \opd{b}_{\vec{k}n} \op{b}_{\vec{k}m} +  \mathcal{B}_{\vec{k}nm} \left( \opd{b}_{\vec{k}n} \opd{b}_{-\vec{k}m} + \text{H.c.} \right) \right] \nonumber \\
    &+ \sum_{\vec{q},\lambda, n} \left( c_{\vec{q}\lambda n}^- \op{b}_{-\vec{q},n} + c_{\vec{q}\lambda n}^+ \opd{b}_{\vec{q},n} \right) \left( \opd{a}_{-\vec{q}\lambda} + \op{a}_{\vec{q}\lambda} \right) .
    \label{eq:Hamil_mag_ph}
\end{align}
The first term represents the bare phonon energy. The second term governs the multi-sublattice magnon system, capturing both the normal exchange ($\mathcal{A}_{\mathbf{k}nm}$) and the anomalous pairing ($\mathcal{B}_{\mathbf{k}nm}$) characteristic of antiferromagnets \cite{10.1063/1.5109132, supplemental_material}.
The third term introduces the magnon-phonon interaction. 
Arising as a relativistic correction that requires spin-orbit coupling \cite{mankovsky_PhysRevLett.129.067202_2022}, its strength is dictated by the coupling vertices:
    $c^\pm_{\vec{q}\lambda n} = \sum_{m,l} \sqrt{{\hbar }/\left({{S_n} M_l \omega_{\vec{q}\lambda}}\right)} e^\mu_{\vec{q} \lambda l} D^\pm_{nml}({\vec{q}}) $.
These vertices project the spatial Fourier transform of the spin-lattice coupling tensor, $\tilde{J}^{\alpha\beta\mu}_{nml}(\vec{q}) = \sum_{jk} J_{in,jm,kl}^{\alpha\beta\gamma} \e{\i \vec{q} \cdot (\vec{r}_i - \vec{r}_k)}$, onto the polarization vectors via the relationship $D^\pm_{nml}({\vec{q}}) = \sigma_m \tilde{J}^{xz,\mu}_{nml}({\vec{q}}) \pm \i \sigma_n\sigma_m \tilde{J}^{yz,\mu}_{nml}({\vec{q}})$, where $\sigma_n = \pm 1$ indicates whether the equilibrium spin on sublattice $n$ is parallel or antiparallel to the quantization axis (see Supplemental Material \cite{supplemental_material}).

Physically, these vertices include on-site and two-site anisotropy and a DMI-like interaction derived from the symmetric and antisymmetric parts of the SLC tensor. While standard DMI is forbidden in centrosymmetric lattices, this dynamic phonon-induced term remains finite. Because conventional macroscopic magnetoelastic theories \cite{Kittel_1949_RevModPhys.21.541} inherently lack this interaction, they are insufficient to capture the hybridization and anomalous transport reported here. 
Notably, this quantitative bulk framework is distinct from minimal 2D altermagnetic toy models \cite{Bendin2026}, where broken out-of-plane mirror symmetry and interfacial DMI are required to generate $d$-wave patterns.

\begin{figure}[t]
    \centering
    \includegraphics[width=1\linewidth]{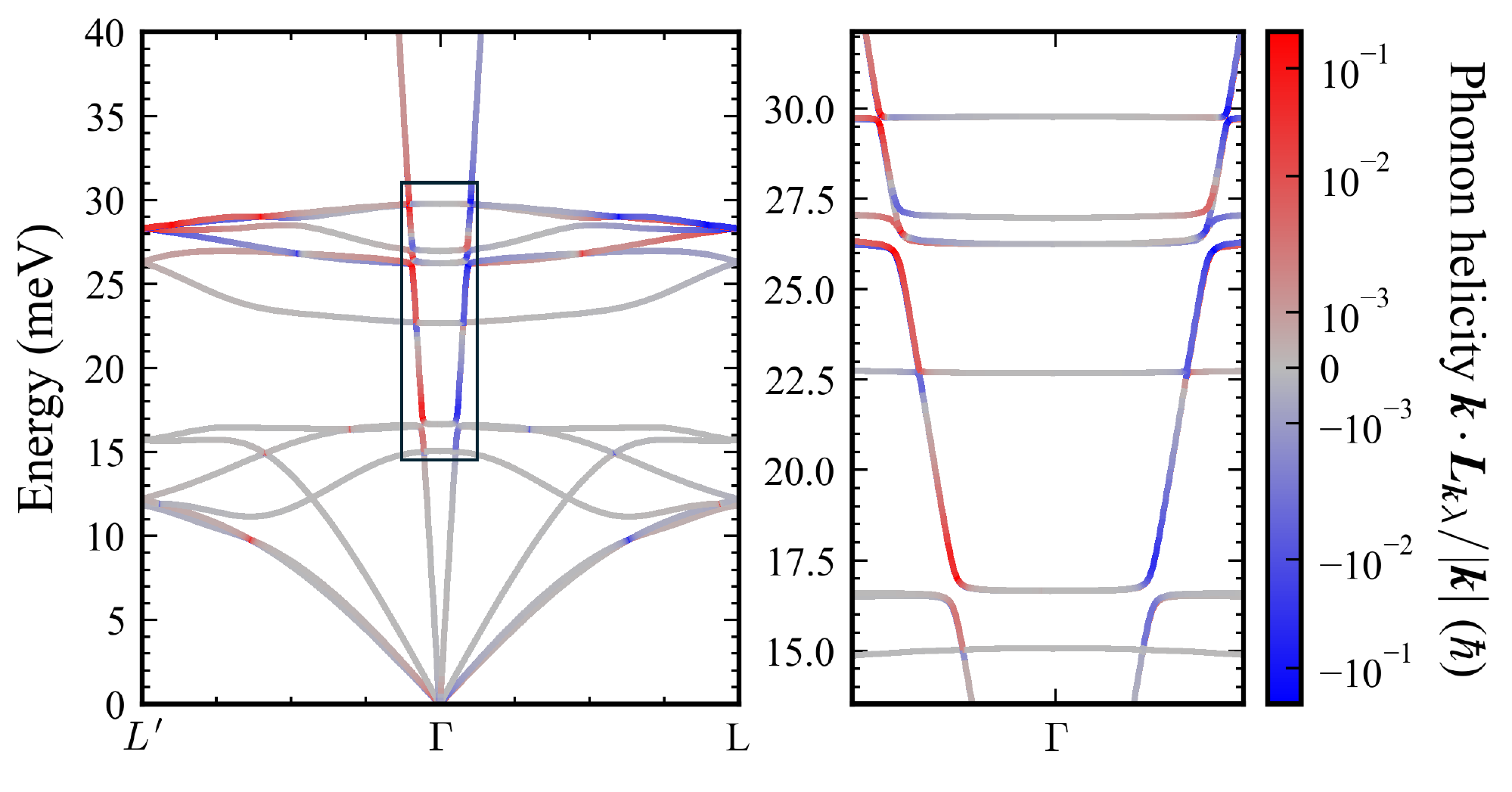}
    \caption{Coupled magnon-phonon bands in CrSb calculated along high symmetry paths, with $\Gamma = (0,0,0)$, $L = (-\pi/a,-\pi/\sqrt{3}a,\pi/c)$ and $L^\prime = (-2\pi/\sqrt{3}a ,0,\pi /c)$. The color scale represents the phonon helicity. The right panel displays a magnified view of the boxed region from the left panel, explicitly showing the avoided crossings (gap openings).}
    \label{fig:hybrid_spectrum}
\end{figure}

\textit{Coupled magnon-phonon spectrum}---
We perform an exact diagonalization of the coupled Hamiltonian (\ref{eq:Hamil_mag_ph}) to find its eigenstates $\ket{n_{\vec{k}}}$ using Colpa's algorithm \cite{colpa_diagonalization_1978, supplemental_material}. Figure~\ref{fig:hybrid_spectrum} displays the resulting energy bands $\varepsilon_{\vec{k}n} = \bra{n_{\vec{k}}}\op{\mathcal{H}}_{\mathrm{mp}}\ket{n_{\vec{k}}}$ along the $\Gamma-L$ and $\Gamma-L^\prime$ paths, along which the bare magnon branches exhibit altermagnetic spin splitting \cite{supplemental_material}. 
The branches with the highest group velocity near the $\Gamma$ point are predominantly magnonic.
At the intersections of the bare magnon and phonon branches, SLC lifts the degeneracies via avoided crossings, opening energy gaps of up to $\SI{1.1}{\milli\electronvolt}$ that signify the formation of magnon-polarons \cite{li_advances_2021}.

The real-space PAM operator is defined as $\hat{\mathbf{L}} = \sum_{il} \hat{\mathbf{u}}_{il} \times \hat{\mathbf{p}}_{il}$, where $\hat{\mathbf{u}}_{il}$ and $\hat{\mathbf{p}}_{il}$ are the displacement and momentum operators of atom $l$ in unit cell $i$. 
In the basis of bare phonons, this yields the expression derived in Ref.~\cite{Zhang_Niu_PRL}:
\begin{align}
\mathbf{L} = \sum_{\mathbf{k},\lambda} \mathbf{L}_{\mathbf{k}{\lambda}} \Big(n_{\mathbf{k}{\lambda}} + \frac{1}{2} \Big) , ~~ \mathbf{L}_{\mathbf{k}\lambda} = i\hbar \sum_{l} \mathbf{e}_{\mathbf{k}\lambda l} \times \mathbf{e}_{\mathbf{k}\lambda l}^* \, , 
\end{align}
where $n_{\mathbf{k}\lambda}$ is the phonon occupation. 
The $1/2$ term dictates that symmetry-permitting systems can host a net PAM at zero temperature due to zero-point fluctuations.

Shifting to the eigenbasis of magnon-polarons, we evaluate the PAM expectation value as $\mathbf{L}_{n,\mathbf{k}} = \langle n_{\mathbf{k}}| \hat{\mathbf{L}} |n_{\mathbf{k}}\rangle$. 
Through hybridization with the altermagnetic magnon modes, the phonons acquire finite PAM. 
Projecting the PAM onto the propagation direction yields the phonon helicity $h_{n,\mathbf{k}} = \mathbf{k} \cdot \mathbf{L}_{n,\mathbf{k}} / |\mathbf{k}|$ (color scale in Figure~\ref{fig:hybrid_spectrum}), a non-zero value of which classifies the phonon as \textit{chiral}  \cite{chiralnature_jurascheck_2025}. The induced helicity is highly anisotropic, reversing sign between the $\Gamma-L$ and $\Gamma-L^\prime$ paths. Because these paths are separated by a $60^\circ$ rotation around the quantization axis, this momentum-space anisotropy motivates a group-theoretical analysis of the PAM.

\begin{figure}
    \centering
    \includegraphics[width=0.99\linewidth]{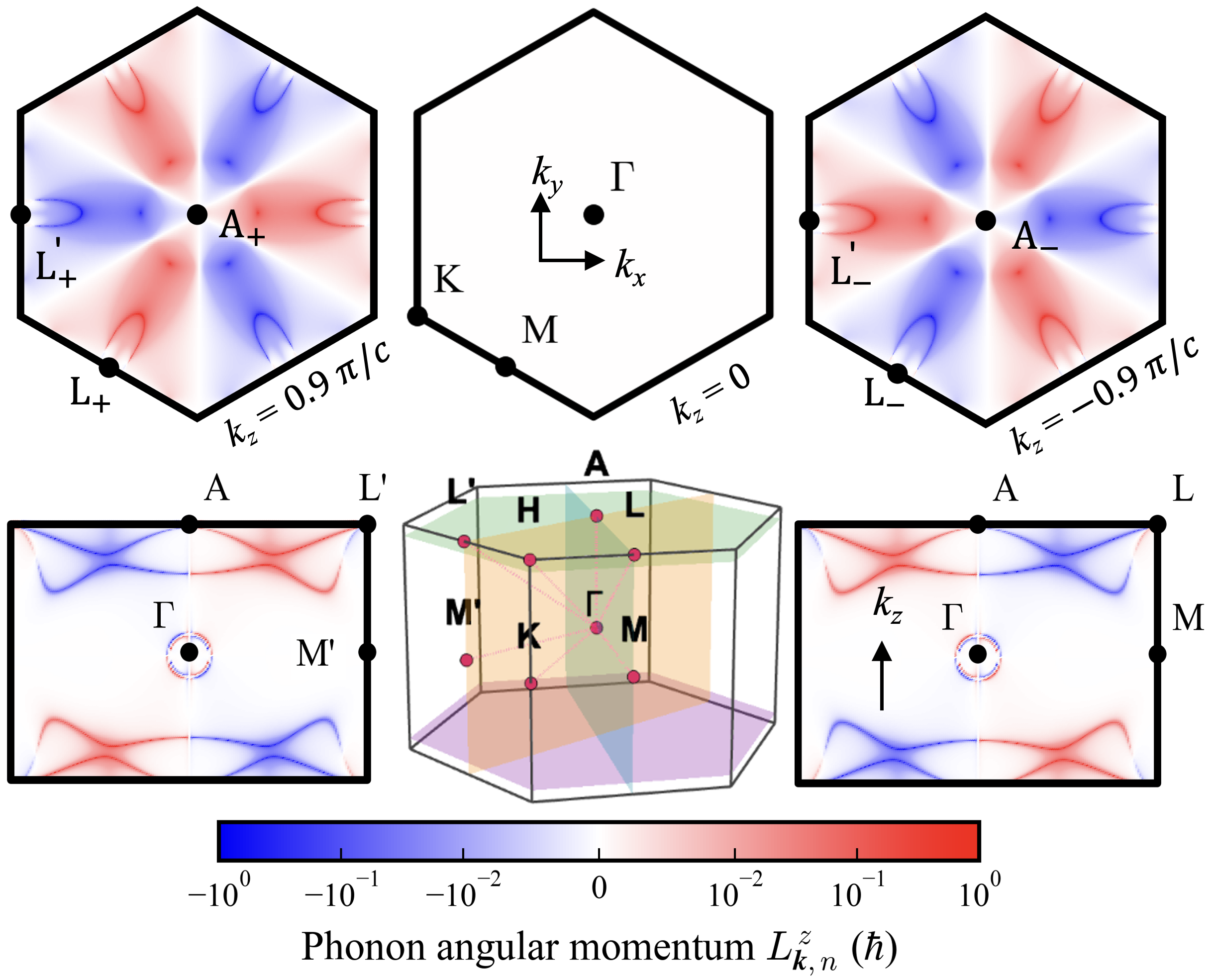}
    \caption{Momentum-space distribution of the PAM for an exemplary hybridized magnon-polaron band in CrSb. Top panels display horizontal slices at $k_z = 0$ and $\pm 0.9 \pi/c$ ($c = 5.352$ \AA), where the shifted high-symmetry points are defined as $L_\pm = L \mp (0.1\pi/c)\mathbf{e}_z$ (analogously for $L^\prime_\pm$ and $A_\pm$). Bottom panels show vertical cuts, with all plane orientations and locations detailed in the central BZ schematic. Bands are indexed by sorting their eigenvalues by energy. Log-scaling { is} applied with linearity below 10$^{-2}\hbar$.}
    \label{fig:PAM_planes}
\end{figure}

\textit{Symmetry-enforced $g$-wave PAM}---
The momentum-space distribution of the PAM is dictated by CrSb's point group $6'/m'mm'$ (Table~\ref{tab:symmetries_combined}). 
Each symmetry operation constrains how $L_{n,\vec{k}}$ must transform between $\vec{k}$-points, and together they enforce a $g$-wave pattern with four nodal planes intersecting at the $\Gamma$ point.
Figure~\ref{fig:PAM_planes} visualizes this highly anisotropic PAM across BZ planes passing through the high-symmetry points $(L,L^\prime,A)$, where the altermagnetic splitting of the bare magnons reaches its maximum \cite{supplemental_material}. 

\begin{table}[t]
\centering
\caption{
Symmetry analysis of momentum-space quantities in CrSb (magnetic point group $6'/m'mm'$), including spin-orbit coupling. The table presents the parity  of the PAM $\vec{L}_{\vec{k}\lambda}$, the phonon helicity $\vec{k} \cdot \vec{L}_{\vec{k}\lambda}$, the normal Berry curvature $\Omega_{xy}$, and the PAM Berry curvature $\Omega_{xy}^{L^z}$ under selected symmetry operations. Group generators are $\{P, C'_{2y}, C'_{6z}, m_z^\prime, C_{3z}\}$, where $m_x = P (C^\prime_{6z})^3 C_{2y}^\prime$ and $C_{2z}^\prime = P m^\prime_z$.}
\label{tab:symmetries_combined}
\renewcommand{\arraystretch}{1.3}
\begin{tabular}{lcccccc}
\hline \hline
{Symmetry operation} & $L^x$ & $L^y$ & $L^z$ & $\vec{k}\cdot\vec{L}$ & $\Omega_{xy}$ & $\Omega_{xy}^{L^z}$ \\ \hline
$m_x$ ($k_x \to -k_x$) & even & odd & odd & odd & odd & even \\
$C'_{2y}$ ($k_y \to -k_y$) & even & odd & even & even & even & even \\
$C'_{2z}$ ($k_z \to -k_z$) & even & even & odd & even & odd & even \\
$m'_z$ ($k_{x,y} \to -k_{x,y}$) & even & even & odd & odd & odd & even \\
$\mathcal{P}$ ($\vec{k} \to -\vec{k}$) & even & even & even & odd & even & even \\ \hline \hline
\label{tab:symm}
\end{tabular}
\end{table}

The time-reversed rotation $C_{2z}^\prime$ ($k_z \rightarrow -k_z$) imposes an odd parity on $L_{n,\vec{k}}^z$, requiring a horizontal nodal plane at $k_z=0$. Similarly, the vertical mirror $m_x$ ($k_x \rightarrow -k_x$) creates a vertical nodal plane at $k_x=0$, which $C_{3z}$ symmetry replicates at $\pm 120^\circ$. The alternating sign of adjacent sectors is dictated by the time-reversed sixfold rotation, $C_{6z}^\prime$. Because $\mathcal{T}$ flips the sign of a pseudovector, the mapping $\mathbf{k} \to C_{6z}\mathbf{k}$ requires that the azimuthal PAM reverses between adjacent sectors ($L^z_{n, C_{6z}\mathbf{k}} = -L^z_{n, \mathbf{k}}$). By mapping the $\Gamma-L$ path into the neighboring $\Gamma-L^\prime$ sector, the $C_{6z}^\prime$ operation enforces a strict reversal of the phonon helicity ($h_{n,\mathbf{k}} \to -h_{n,\mathbf{k}}$). 
This explains the anisotropic, opposite-chirality band splittings observed in the magnon-polaron spectrum in Fig.~\ref{fig:hybrid_spectrum}, establishing that bulk CrSb hosts a $g$-wave PAM distribution arising from chirality-selective hybridization.

\begin{figure}[]
    \centering
    \vspace*{-0.3cm} 
\includegraphics[width=0.7\linewidth]{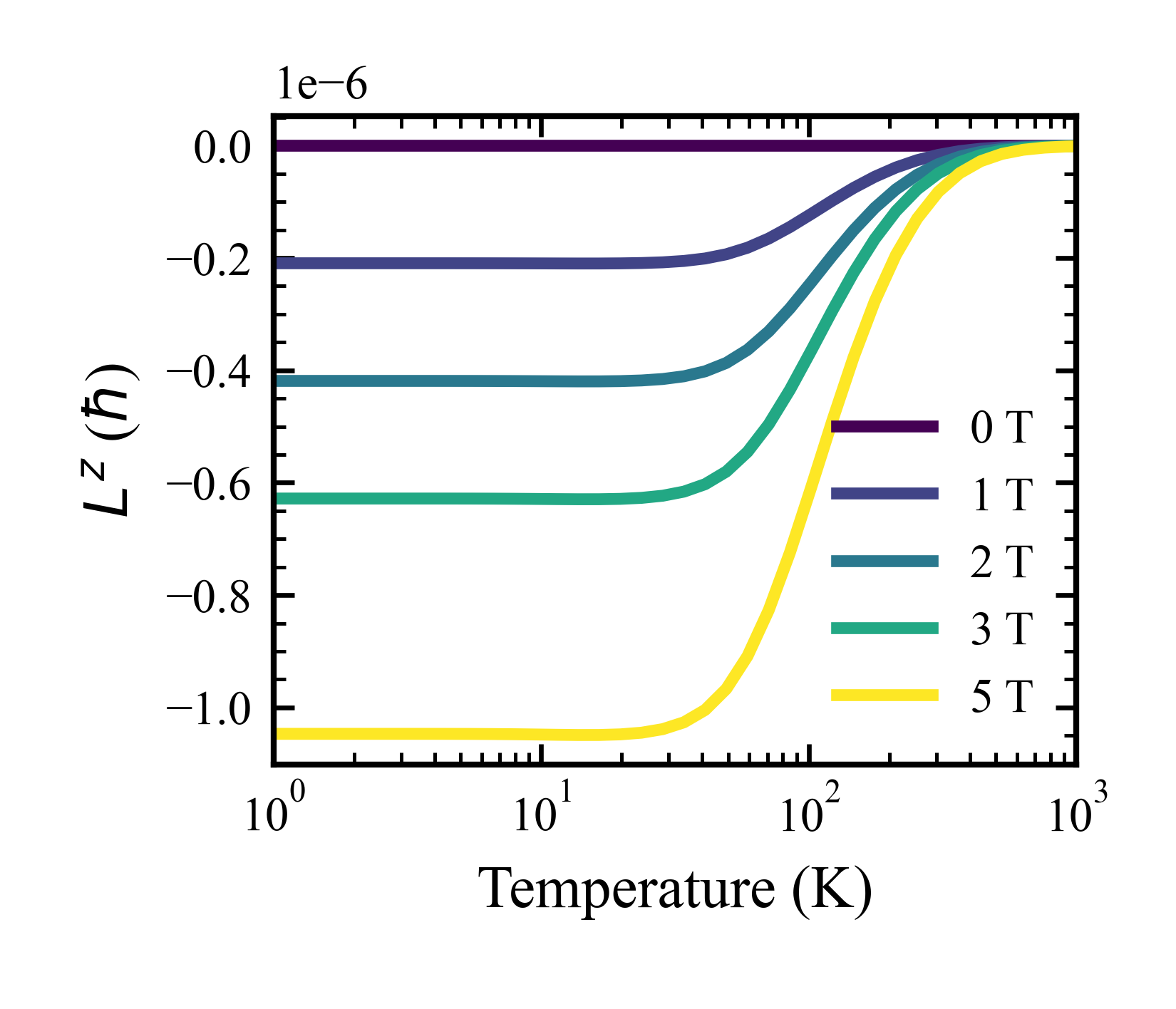}
    \vspace*{-0.6cm}
    \caption{Temperature dependence of the equilibrium phonon angular momentum $L^z$ per unit cell in CrSb, calculated for various applied magnetic fields.}
    \label{fig:zero_point_PAM}
\end{figure}

Next, we compute the total PAM. 
Although individual chiral phonons carry PAM, the total equilibrium PAM vanishes identically in the field-free state (Fig.~\ref{fig:zero_point_PAM}). This global compensation is enforced by the $6'/m'mm'$ point group: the out-of-plane component $L^z$ cancels pairwise across the  BZ due to its odd transformation under $m_x$ and $C'_{2z}$ (Table~\ref{tab:symmetries_combined}), while any net in-plane PAM ($L^x, L^y$) is strictly forbidden by the $C_{3z}$ rotational symmetry. By having momentum-dependent chiral splittings while maintaining a compensated total angular momentum, the lattice establishes a direct phononic analogue to altermagnetism.
Applying an external field along the $z$-axis breaks $m_x$ and $C'_{2z}$, lowering the magnetic point group to $\bar{3}m^\prime$ and preventing pairwise energetic cancellation, resulting in a finite zero-point PAM, $L^z$ (Fig.~\ref{fig:zero_point_PAM}). At elevated temperatures, $L^z$ vanishes asymptotically \cite{Markus_2025_PRL, Zhang_Niu_PRL}. 

\textit{Anomalous Nernst transport}---
Anomalous transverse transport of heat \cite{zhang_thermal_2019, bao_direct_2023, Ma_2024_PhysRevLett.133.246604} and spin \cite{Park_2_2020_doi:10.1021/acs.nanolett.0c00363, Klogetvedt_2023_PhysRevB.108.224424, Markus_2025_PRL} driven by magnon-polaron formation has thus far been demonstrated in systems possessing net magnetization or subject to external magnetic fields — including ferromagnets, multiferroics, and field-polarized antiferromagnets.
By extending purely magnonic \cite{Cui_PhysRevB.108.L180401} and phononic \cite{Park_2020_doi:10.1021/acs.nanolett.0c03220} transport models to the hybridized spin-lattice regime, we demonstrate how altermagnets activate anomalous transport without net magnetization or an applied field.

{The linear transport coefficients are given as $J^\mathcal{O}_\mu = -\sum_{\nu} \alpha^\mathcal{O}_{\mu \nu} \nabla_\nu T$, where the operator $\hat{\mathcal{O}} \in 
\{2g^{-1}, \hbar\hat{S}^\tau, \hat{L}^\tau \}$ selects the normal, spin, and PAM transport, respectively.}
The $\alpha^\mathcal{O}_{\mu \nu}$ coefficients ($\mu \neq \nu$) are related to the generalized Berry curvature, defined as
\begin{equation}\Omega_{n, \mathcal{O}}^{\mu\nu}(\mathbf{k}) = 2i\hbar^2 \sum_{m \neq n}^{2N_b} \frac{g_{nn} g_{mm} \langle n_\mathbf{k}| \hat{j}_\mu^{\mathcal{O}} | m_\mathbf{k} \rangle \langle m_\mathbf{k}| \hat{v}_\nu | n_\mathbf{k} \rangle}{[(g\mathcal{E}_\mathbf{k})_{nn} - (g\mathcal{E}_\mathbf{k})_{mm}]^2}  .
\label{eq:BC}
\end{equation}
Here $\hat{j}_\mu^{\mathcal{O}}$ is the generalized current operator, $\hat{j}_\mu^{\mathcal{O}} = \frac{1}{4} \{ \hat{v}_\mu, g\hat{\mathcal{O}}\}$, $\hat{v}_\nu = \partial \hat{\mathcal{H}} / \partial (\hbar k_\nu)$ is the velocity operator, and $g = \sigma_z \otimes \mathbf{I}_{N_b \times N_b}$, with $\sigma_z$ being the third Pauli matrix and $\mathbf{I}_{N_b \times N_b}$ the identity matrix for the $N_b$ bands, and $\mathcal{E}_{\mathbf{k}}=\text{diag}(\varepsilon_{\mathbf{k}1},\dots,\varepsilon_{\mathbf{k}N_b},\varepsilon_{-\mathbf{k}1},\dots , \varepsilon_{-\mathbf{k}N_b})$ is the diagonal matrix of eigenenergies. 
Integrating the spin and PAM curvature over the BZ yields the macroscopic anomalous Nernst coefficients for spin and PAM:
\begin{equation}
\alpha_{\mu\nu}^{\mathcal{O}} = -\frac{2k_B}{V} \sum_{\mathbf{k}} \sum_{n=1}^{N_b} c_1(n_{\mathbf{k}n}) \Omega_{n, \mathcal{O}}^{\mu\nu}(\mathbf{k}) ,
\end{equation}
where $c_1(x) = (1+x)\ln(1+x) - x\ln x$ is the weighting function, $V$ is the system's volume, and $n_{\mathbf{k}n}$ is the Bose-Einstein distribution.

\begin{figure}
    \centering
    \includegraphics[width=0.90\linewidth]{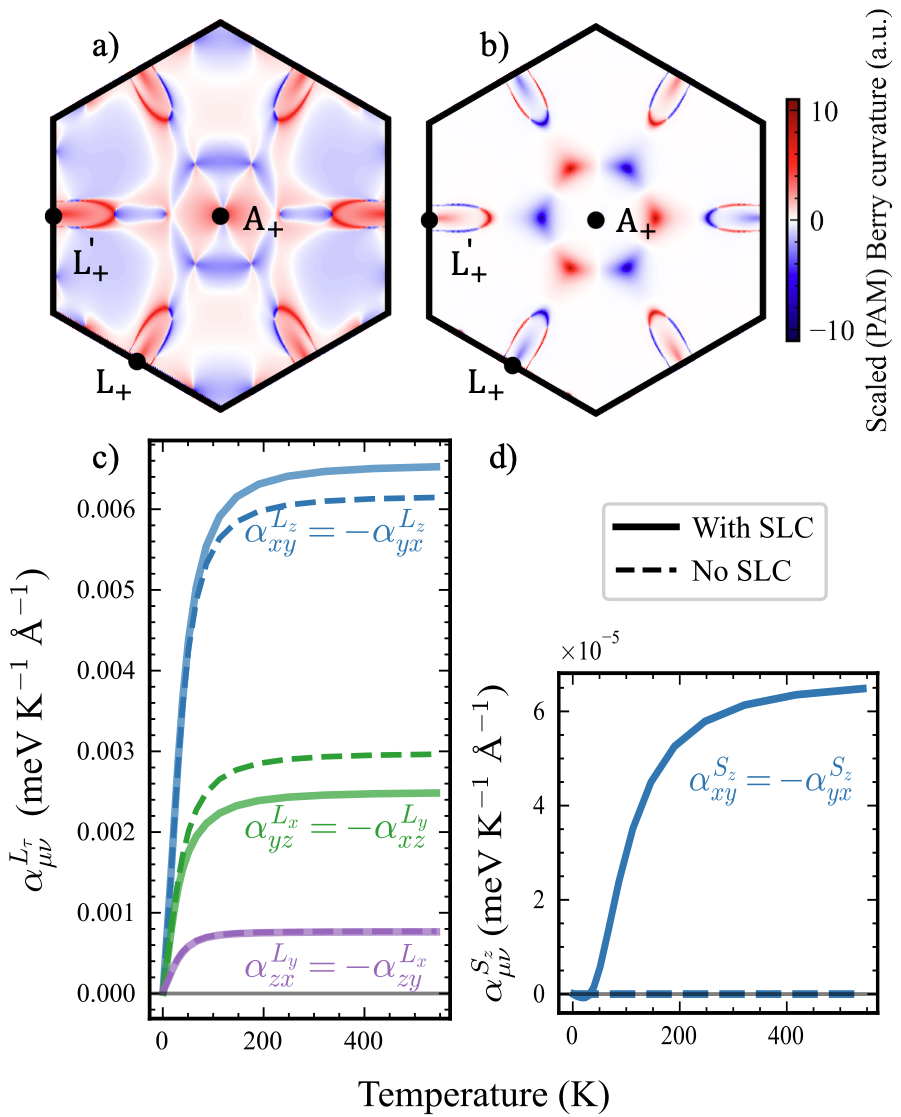}
    \caption{Berry curvatures and resulting anomalous transport of magnon-polarons in bulk CrSb. (a)  PAM $\Omega_{xy}^{L^z}$ and (b) normal $\Omega_{xy}$ Berry curvatures for a representative hybridized band evaluated in the $k_z = 0.9\pi/c$ plane, where the shifted high-symmetry points are defined as in Fig. \ref{fig:PAM_planes}. A logarithmic color scale, $\text{sgn}(\Omega)\ln(1+ |\Omega|)$, is applied. Calculated anomalous (c) PAM and (d) spin Nernst coefficients. Solid (dashed) curves correspond to calculations with (without) SLC. Symmetry-enforced, non-vanishing tensor components are labeled.}
    \label{fig:nernst_coefficients}
\end{figure}

Figures \ref{fig:nernst_coefficients}(a),(b) show the computed PAM and normal Berry curvatures, $\Omega_{xy}^{L^z}$ and $\Omega_{xy}$, respectively, for the  $k_z= 0.9\pi/c$ plane. These obey the symmetries detailed in Table \ref{tab:symm}. 
The computed Nernst tensors [Figs.~\ref{fig:nernst_coefficients}(c),(d)] reproduce the symmetry-imposed form tabulated for this magnetic point group in linear-response theory \cite{Seemann_2015_PhysRevB.92.155138}.
For PAM transport [Fig.\ \ref{fig:nernst_coefficients}(c)], the anomalous Nernst coefficient exhibits a finite background even in the absence of SLC. At low temperatures, this signal grows linearly, driven exclusively by acoustic modes. This aligns with theoretical predictions for the universal, nonrelativistic PAM {thermal} Hall effect \cite{Park_2020_doi:10.1021/acs.nanolett.0c03220}. 
However, introducing relativistic SLC modifies the transport at elevated temperatures. 
Because the hybridization gaps open at higher energies (starting around \SI{11}{\milli\electronvolt}), these regions of significant Berry curvature remain unoccupied for $T \approx 0$.
Once the thermal energy crosses this threshold (above \SI{100}{\kelvin}), 
the population of hybridized states modifies the PAM Nernst transport: 
the components $\alpha^{L^z}_{xy} = -\alpha^{L^z}_{yx}$ are enhanced, 
while $\alpha^{L^x}_{yz} = -\alpha^{L^y}_{xz}$ are reduced 
relative to the nonrelativistic background.

In contrast, the anomalous spin Nernst coefficient [Fig.\ \ref{fig:nernst_coefficients}(d)] vanishes identically without spin-orbit interactions, providing an intrinsic, magnetically driven counterpart to spin currents activated by structural chirality \cite{kim_chiral-phonon-activated_2023} or the nonrelativistic electrical spin-splitter effect \cite{PhysRevLett.126.127701}.

We note that while transverse spin and PAM currents are symmetry-allowed, the zero-field anomalous thermal Hall effect vanishes. The normal Berry curvature $\Omega_{n}(\mathbf{k})$ is odd under at least one momentum-space symmetry (see Fig.~\ref{fig:nernst_coefficients}(b), Table \ref{tab:symmetries_combined}). Upon integration over the full BZ, the odd Berry curvature cancels exactly, yielding a zero net anomalous transverse thermal conductivity.
In bulk CrSb, the magnitude of the PAM Nernst coefficient exceeds the pure spin Nernst effect by two orders of magnitude, with relativistic SLC contributing up to $\sim 20\%$ of the total PAM signal at room temperature.

Lastly, we  note that, within linear response theory, the macroscopic  spin and PAM currents $J^\mathcal{O}_\mu \propto \nabla_\nu T$  are periodic in the angle of the temperature gradient $\nabla T$ with period $\pi$ at most (see SM \cite{supplemental_material}).
The $g$-wave symmetry of the PAM distribution is therefore not reflected in the linear-response current; resolving it would require accessing the nonlinear transport regime.

\textit{Conclusion}--- 
We have demonstrated the existence of chiral phonons at zero magnetic field arising from the hybridization of altermagnetic magnons with phonons. 
Relativistic SLC imprints the unconventional symmetry of the altermagnetic phase onto
the phonon angular momentum distribution across the full BZ, as we have shown exemplary using first-principles calculations for the bulk altermagnet CrSb.
Within these hybridized magnon-phonon states, the PAM acquires a characteristic $g$-wave distribution, unlike conventional collinear antiferromagnets, where $\mathcal{PT}$ symmetry forces the PAM to vanish at every $\mathbf{k}$-point and suppresses Berry curvature throughout the BZ.
This PAM distribution establishes a direct phononic analogue to the compensated spin-split bands of altermagnets, by lifting the degeneracy of opposite chirality modes while enforcing an exact cancellation across the BZ. 
We emphasize that this mechanism is distinct from the recently proposed PAM induced by electron-phonon coupling \cite{Wang2025,wang2025alteraxial}.
We further revealed that these magnon-polarons exhibit finite Berry curvatures that drive anomalous transverse transport. 
While the PAM Nernst response features a nonrelativistic background modified by this hybridization, the macroscopic spin Nernst effect is activated entirely by the hybridization. 
This work establishes bulk altermagnets as a suitable materials class for the generation of chiral phonons and transport -- requiring neither net magnetization, applied magnetic field, nor structural chirality -- that will be beneficial for spin caloritronics and chiral phononics.

\textit{Acknowledgments}---
We acknowledge support by the Swedish Research Council (VR), the German Research Foundation (Deutsche Forschungsgemeinschaft) through CRC/TRR 227 “Ultrafast Spin Dynamics” (project MF, Project ID 328545488), and
the K.\ and A.\ Wallenberg Foundation (Grants No.\ 2022.0079 and No.\ 2023.0336). Part of the calculations were supported by the National Academic Infrastructure for Supercomputing in Sweden (NAISS) at NSC Link\"oping, funded by VR through Grant No.\ 2022-06725.

\textit{Data availability}---
The data that support the findings of
this article are available from the authors upon reasonable
request.

\bibliography{references}

\end{document}